\documentclass[traditabstract]{aa}

\usepackage{graphicx}
\usepackage{amssymb,amsmath}
\usepackage{txfonts}
\usepackage{natbib}
\usepackage{rotating,sidecap}

\begin{document}

   \title{X-ray and UV investigation into the magnetic connectivity of a solar flare}

   \author{H. A. S. Reid, N. Vilmer, G. Aulanier and E. Pariat}
   \institute{LESIA, Observatoire de Paris, CNRS, UPMC, Universit\'{e} Paris-Diderot, 5 place Jules Janssen, 92195 Meudon Cedex, France}

   \date{Received / Accepted }

\abstract{We investigate the X-ray and UV emission detected by RHESSI and TRACE in the context of a solar flare on the 16th November 2002 with the goal of better understanding the evolution of the flare.  We analysed the characteristics of the X-ray emission in the 12-25 and 25-50 keV energy range while we looked at the UV emission at 1600 $\text{\AA}$.  The flare appears to have two distinct phases of emission separated by a 25-second time delay, with the first phase being energetically more important.  We found good temporal and spatial agreement between the 25-50 keV X-rays and the most intense areas of the 1600 $\text{\AA}$ UV emission.  We also observed an extended 100-arcsecond $<$ 25 keV source that appears coronal in nature and connects two separated UV ribbons later in the flare.  Using the observational properties in X-ray and UV wavelengths, we propose two explanations for the flare evolution in relation to the spine/fan magnetic field topology and the accelerated electrons.  We find that a combination of quasi separatrix layer reconnection and null-point reconnection is required to account for the observed properties of the X-ray and UV emission.}



\titlerunning{X-ray and UV Emission of a Solar Flare}
\authorrunning{Reid et al}

\maketitle

\section{Introduction}

Solar flares are known to produce non-thermal populations of electrons in the corona, which stream down the magnetic field to produce UV and X-ray emission in the upper chromosphere \citep[see][for an observational review]{Fletcher_etal2011}.  In this scenario the geometry of the magnetic field plays a very important role in determining where electrons can travel and where their subsequent electromagnetic signature will be emitted.  Using X-ray observations from RHESSI \citep{Lin_etal2002}, we investigated this question in a confined flare that was previously studied by \citet{Masson_etal2009, Baumann_etal2012}, who analysed the coronal magnetic field with respect to UV observations detected by TRACE \citep{Handy_etal1998}.

An early comparison between extreme ultraviolet (EUV) and HXR (hard X-ray) emissions in solar flares found a good temporal agreement \citep{KaneDonnelly1971,DonnellyKane1978,Kane_etal1979}.  The HXR emission analysed was mainly in the deca-keV range while the EUV spanned the wavelengths between 10 and 1030~$\text{\AA}$.  The peak flux of the HXR and EUV lightcurves exhibited the strongest correlation.  The duration of rise and decay phases tended to be longer for EUV emission, with the closest agreement coming from the lower energy X-ray photons around $10$~keV.  A better temporal agreement for rise times has been found in many subsequent studies which concentrated on UV emission between 1400-1600~$\text{\AA}$ and $>25$~keV X-rays \citep[e.g.][]{Cheng_etal1981,Woodgate_etal1983,Cheng_etal1988,WarrenWarshall2001,AlexanderCoyner2006,CoynerAlexander2009}.  The good temporal correlation indicates a common non-thermal electron source.  Recent modelling of this emission process has been performed with a 1D approach that incorporates radiative transfer and hydrodynamics \citep{AbbettHawley1999,Allred_etal2005}.  Non-thermal electrons propagating into the lower atmosphere were found to produce optical line and continuum emission similar to what is observed during solar flares.


Good spatial agreement has also been found between UV and HXR emission.  \citet{WarrenWarshall2001} found that HXR-UV co-temporal emission tended to be co-spatial and vice versa in a collection of about ten flares.  \citet{AlexanderCoyner2006,CoynerAlexander2009} found that HXR footpoints are smaller than their correlated UV ribbons, only overlapping with a part of the UV ribbon.  A counter example has been detected \citep{Liu_etal2007,Liu_etal2008} in which HXR sources were detected all along a UV ribbon, but such examples are rare and are possibly due to a more uniform energy release rate along the ribbon.  Lower energy X-rays ($6-25$~keV) have been observed \citep{AlexanderCoyner2006,Liu_etal2007} later in flares to connect two separate UV ribbons which implies magnetic connectivity between ribbons in the corona.  The same energy range of X-rays was found to temporally correlate to late-phase UV emission in the flare on the 6th December 2006 \citep{CoynerAlexander2009} in which the HXRs were \emph{not} spatially correlated and only weakly co-temporal.

In addition, results from UV and HXR studies during flares have highlighted the close connection between the 3D geometry and topology of the overlying magnetic field and the spatial distribution of the UV and HXR emission.  Observations have found that H$\alpha$ and UV ribbons are located at the footpoints of particular field lines such as separatrices  \citep[e.g.][]{GorbachevSomov1988,Mandrini_etal1991} or quasi-separatrix layers \citep[QSLs,][]{Demoulin_etal1996, Demoulin_etal1997,Chandra_etal2011}.  These QSLs are regions of very strong magnetic connectivity gradients (and by extension regions of magnetic connectivity discontinuities, separatrices) and define preferential sites for the build-up of intense electric current sheets involved in the magnetic reconnection process.  Particles accelerated from the reconnection region can flow along these separatrices/QSLs and impact the lowest layer of the atmosphere, hence explaining the correlation between the UV and HXR emission to the 3D topological structure of the magnetic field.  Hard X-ray footpoints have been observed in which reconnection and energy release rates are highest along flare ribbons \citep{Asai_etal2002,Asai_etal2004,Temmer_etal2007}.  This scenario implies that we require a much higher energy density of electrons to produce HXRs compared to the UV flare ribbons.


Much work has gone into understanding the 3D topological structures of coronal magnetic fields and how reconnection can occur \citep[e.g.][and references therein]{PriestForbes2002}. Reconnection at separatrix field lines, such as the fan and spine separatrices passing by a 3D magnetic null point \citep{LauFinn1990}, involves a strict one-to-one reconnection of magnetic field lines \citep[e.g.][and references therein]{Pontin_etal2007,Pariat_etal2009}. Quasi-separatrix layer reconnection, named slip-running reconnection in its super-Alfv\'{e}nic regime \citep{Aulanier_etal2006}, does not conserve field lines: magnetic connectivity is continuously exchanged between neighbouring field lines that are involved in QSL reconnection, and at large scales slip-running field lines are observed to slide relative to each other \citep{Aulanier_etal2006,Aulanier_etal2007,Masson_etal2012}. Depending on the magnetic topology, different reconnection modes can therefore be triggered and lead to different dynamics of accelerated particles.  Observations of UV and HXR emission during flares, linked with the 3D magnetic topology, is therefore helpful for understanding magnetic reconnection processes in the solar atmosphere.


Because of the close relationship between X-ray and UV emission, we analysed these wavelengths from the flare on the 16th November 2002 in the context of magnetic field modelling.  The layout of this article is as follows.  In Section 2 we summarised results from \citet[][hereafter MPAS09]{Masson_etal2009} regarding the magnetic field.  In Section 3 we review both the UV and the HXR spatial and temporal information.  In Section 4 we examine the co-temporal and co-spatial nature of the two wavelength ranges.  We conclude in Section 5 with a discussion of the observations and an interpretation of the flare.

\section{Magnetic environment}

\begin{figure*}\centering
\sidecaption 
\includegraphics[width=12cm]{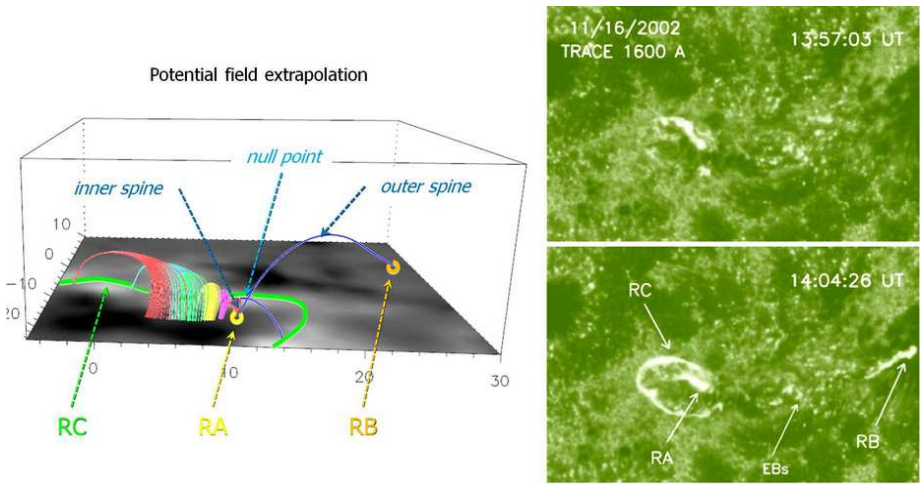}
\caption{Left: potential field extrapolation showing the fan/spine configuration and the coronal null point.  The UV ribbons are also indicated.  Right: a TRACE 1600 $\text{\AA}$ image before and after the flare impulsive phase.  The UV ribbons related to the inner and outer spine field are denoted as RA and RB, respectively.  The quasi-circular UV ribbon related to the fan surface is denoted as RC.  Adapted from MPAS09.}
\label{fig:PF_UV}
\end{figure*}

The focus of the MPAS09 paper was examining a well-defined magnetic topology and how it evolved during a flare (the flare on the 16th November 2002).  A potential field extrapolation (Figure \ref{fig:PF_UV}) found the magnetic field to include a single coronal null point.  The magnetic field took the form of a 3D spine/fan configuration \citep{LauFinn1990}.  The fan surface split the region into two separate domains of magnetic connectivity, which both included a spine separatrix field line.  The two fan eigenvectors that define the fan shape were aligned along the north-south and east-west direction.  The north-south eigenvector is larger by about a factor of 8, producing asymmetry in the fan surface.  

Using the potential field extrapolation as their initial condition, MPAS09 modelled the evolution of the magnetic field with an MHD simulation.  Shearing motions were used to drive the system out of equilibrium and stimulate magnetic reconnection.  An intense current sheet was found to occur at and around the magnetic null point and was caused by a shearing of the two spines in the plane of the fan.  \citet{Masson_etal2009} also found that traditional null point reconnection could explain neither the elongated nature of the flare ribbons associated with the spine nor the sequential illumination brightening of the fan-related ribbon.  Quasi-separatrix layers surrounding both the spine and fan field calculated using the squashing degree \citep{Titov_etal2002} matched the locations of the UV ribbons.  Moreover, slipping and slip-running reconnection induced an apparent motion of field lines and were able to describe in detail the ribbon connected to the outer spine.  This motion of field lines caused by slip-running reconnection has been suggested as a possible explanation for observed HXR footpoint movement during flares \citep{Aulanier_etal2006}.



\section{X-ray and UV data}

\subsection{UV morphology} 

The study of MPAS09 analysed the UV morphology (Figure \ref{fig:PF_UV}) with respect to a potential field extrapolation.  The authors divided the TRACE UV emission into three subregions that were related to the overlying magnetic field.  These were
\begin{enumerate}
\item A quasi-circular ribbon on the left-hand side of the active region related to the fan structure of the magnetic field, RC.
\item A small straight ribbon at the top right of the left hand side quasi-circular ribbon related to the inner spine of the magnetic field, RA.
\item A small straight ribbon on the right-hand side of the active region related to the outer spine of the magnetic field, RB.
\end{enumerate}


A good correlation between the magnetic field spine/fan topology and the TRACE UV emission is expected in the standard flare scenario in which non-thermal electrons stream down the magnetic field and deposit their energy in the upper chromosphere.  The magnetic correlation continues with the QSLs, which are also spatially correlated with the UV ribbons.  We also note that the majority of the emission on the left side of the active region comes from the small straight ribbon (RA) and the top right of the quasi-circular ribbon (RC).  This region is spatially very close to the magnetic null point where a current sheet can form and reconnection can occur \citep[e.g.][]{PriestTitov1996}.  Accelerated electrons would be preferentially directed along the largest eigenvector associated with the null point in the fan of the magnetic field (in this case north-south), but we do not see intense emission south of the magnetic null.  This is perhaps due to the increased divergence of the magnetic field to the south because the null point is closer to the north of the quasi-circular ribbon (RC).  The small straight ribbon on the right side of the active region (RB) is not exactly co-spatial with the outer spine of the potential field extrapolation.  However, MPAS09 accounted for this discrepancy by evoking the time difference of 6.5 hours between the flare and the extrapolation, the non-potentiality in the real flaring region, and the sensitive nature of the outer spine to the extrapolation variables.  The authors postulated that the `real' outer spine must intersect the UV ribbon on the right side of the active region.





\subsection{UV lightcurves}

\begin{figure}\centering
\includegraphics[width=0.99\columnwidth]{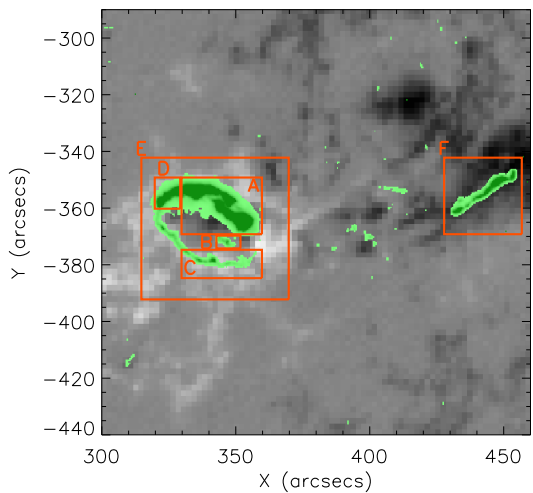}
\caption{TRACE 1600 $\text{\AA}$ emission (green) for the active region on the 16th November 2002 at 13:58:29 UT.  The background is the SOHO MDI magnetogram at 14:24 UT.  The two larger areas E and F enclose the emission on the left-hand and right-hand side of the active region.  The smaller areas A, B, C, and D represent substructure within area E that varies differently in time.  In Figure \ref{fig:hsi_tra_lc} area A, C, and E are normalised to 1 and area B, D, and F are normalised to 0.5.}
\label{fig:UV_morph}
\end{figure}
\begin{figure}\centering
 \includegraphics[width=0.99\columnwidth]{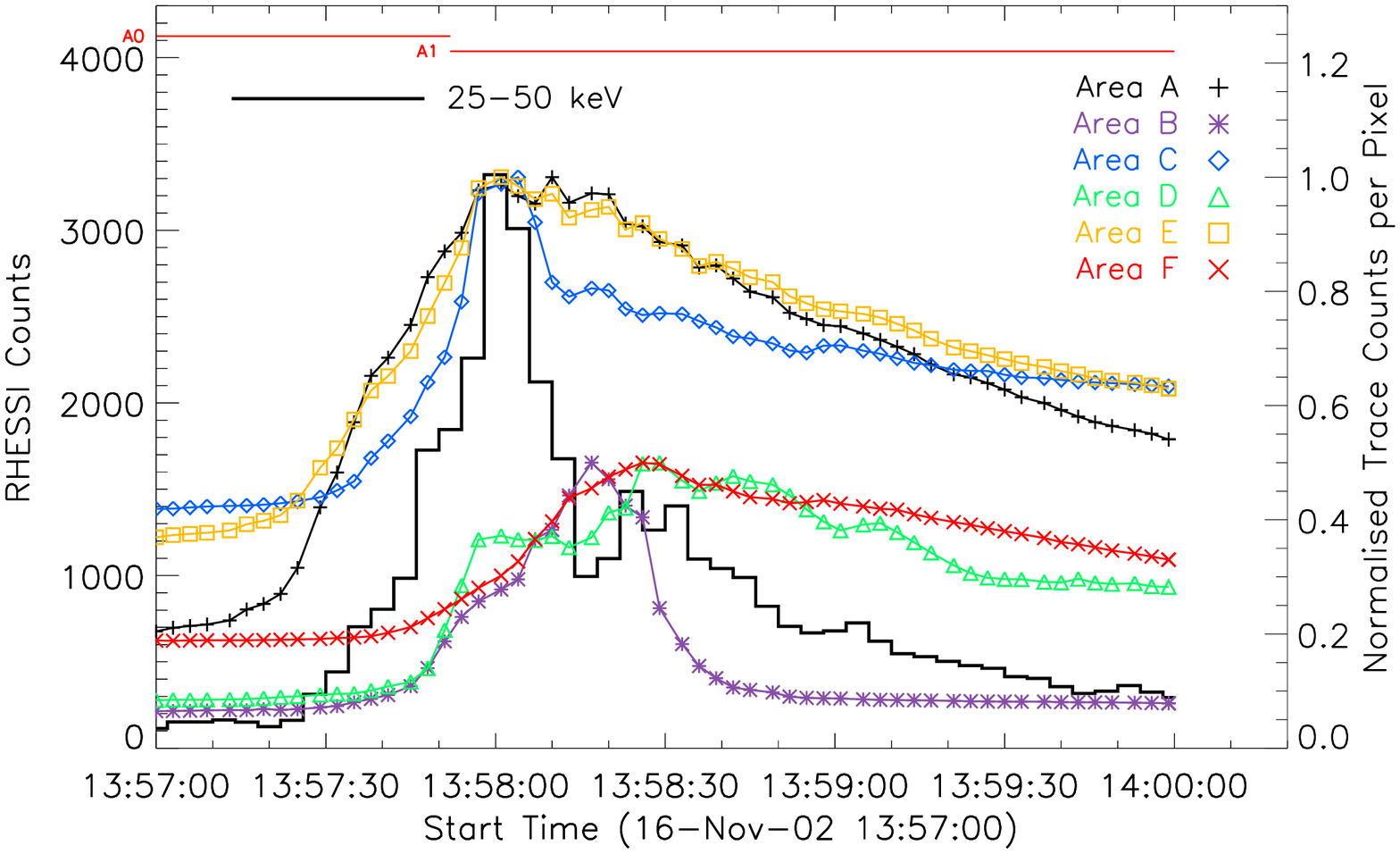}
 \includegraphics[width=0.99\columnwidth]{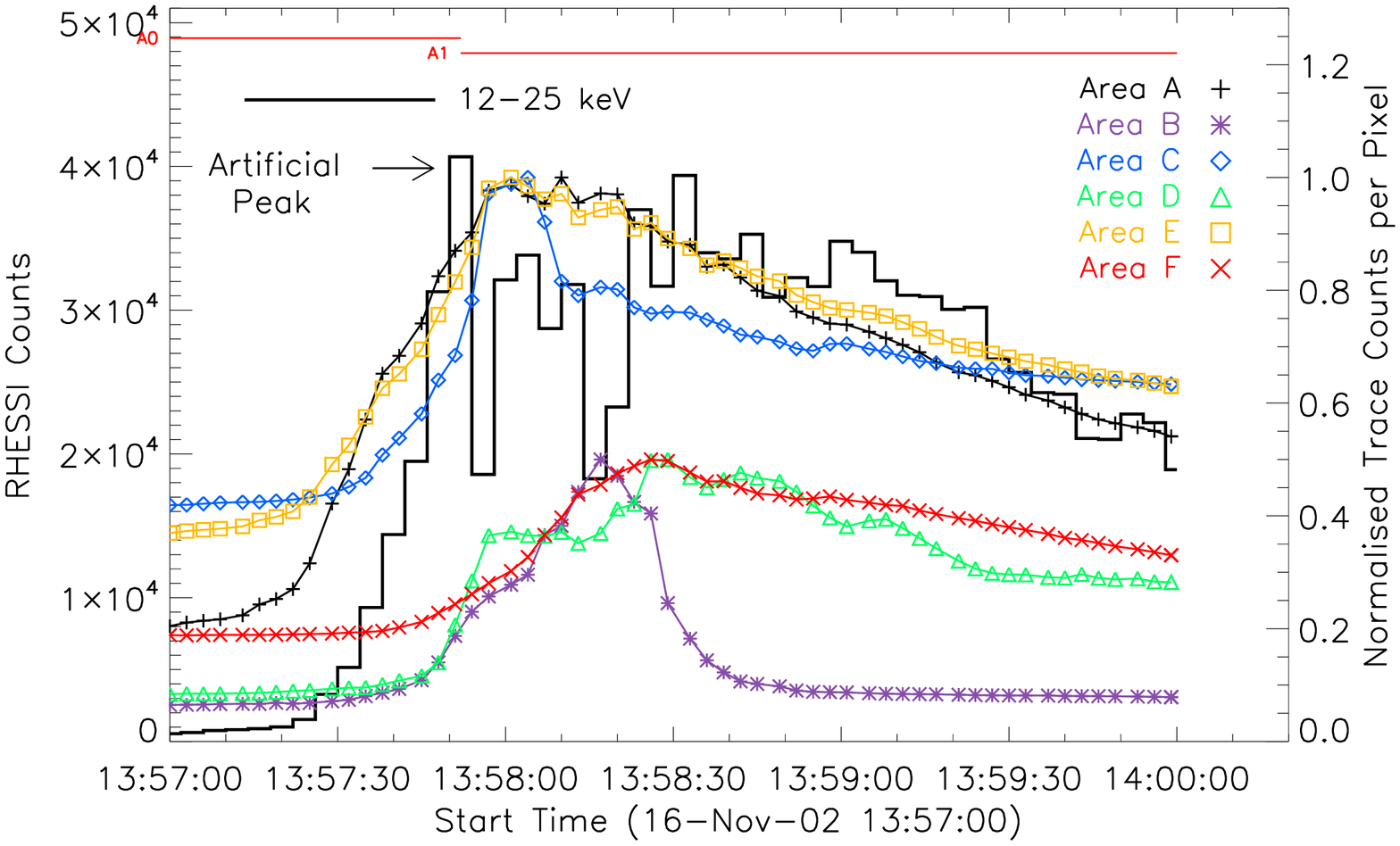}
\caption{Lightcurves from the TRACE 1600 $\text{\AA}$ UV regions in Figure \ref{fig:UV_morph} and the RHESSI lightcurves.  The yellow/red trace curves (area E, and F) in both graphs correspond to the normalised pixel-averaged counts of the left/right regions in Figure \ref{fig:UV_morph}.  Areas A, C, and F are normalised to 1 while areas B, D, and E are normalised to 0.5 for clarity.  Top: the RHESSI lightcurve for the 25-50 keV energy range in counts.  Bottom: the RHESSI lightcurve for the 12-25 keV energy range in counts.  Both RHESSI curves use a time resolution of 4 seconds and all nine detectors.  The change in attenuator state from A0 to A1 is also shown in red at the top of both graphs.  The change in attenuator creates an artificial peak in the 12-25 keV energy range.}
\label{fig:hsi_tra_lc}
\end{figure}

Because we aim to examine how the UV emission varies in time, we examined the substructure in the ribbons defined by MPAS09.  The different areas shown in Figure \ref{fig:UV_morph} show regions of the UV emission that behave differently in time.  Their lightcurves are shown in Figure \ref{fig:hsi_tra_lc}.  We have split the emission on the left-hand side of the active region into subregions (denoted area A, B, C, and D) displayed in Figure \ref{fig:UV_morph}.  The time profiles of the whole UV emission on the left-hand and right-hand side are represented as areas E and F, respectively.  The magnitude of the pixel-averaged flux strongly depends upon the size of the area around the emitting region.  If there are many pixels in the area that does not emit UV radiation, the average flux will be lower.  Moreover, the UV emission saturates the TRACE detector during the brightest periods of emission.  We therefore plot the normalised magnitudes of the lightcurves, where curves A, C, and E are normalised to 1 and curves B, D, and F are normalised to 0.5.  Owing to the saturation of the TRACE detectors, the peak flux will contain some uncertainty.  The peak flux time for curves A, C, and E is approximately 13:58:00 UT which is 25 seconds before the peak flux time for curve D and F around 13:58:25 UT.  We can have more confidence in the delay between peak flux times because a similar time delay is observed between the start time of emission in curves A, C, and E and curves D and F.  Curve B in Figure \ref{fig:UV_morph} peaks by itself at 13:58:20 UT.


Lightcurves A and E are very similar but with a slight deviation around 14:00:00 UT.  This is because the majority of UV emission on the entire left-hand side (curve E) comes from area A.  We note that area A demonstrates the similarity in the time profile of both the inner ribbon RA and the most intense part of the circular ribbon RC.  Curves D and F are similar in their peak and decay profile but curve D rises much faster (around the peak time of curve E) to approximately 2/3 of its peak flux magnitude.  A time correlation in the peak flux of curves D and F suggests that the emission has a common driver, reinforcing the link between the UV emission on the left-hand and right-hand sides.  We also remark on the very fast decay time of curve B, which does not have any features around 13:58:25 UT.

\subsection{X-ray morphology}

\begin{figure*}\centering\sidecaption
 \includegraphics[width=12cm]{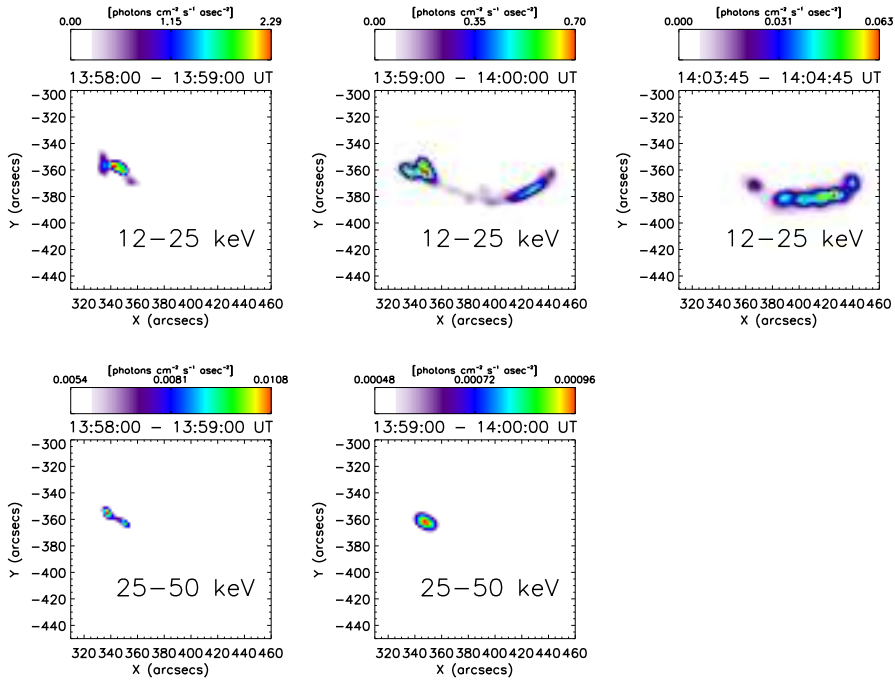} 
\caption{Top: RHESSI pixon images between 12-25 keV using attenuators 3-9.  Bottom: RHESSI pixon images between 25-50 keV using attenuators 3-9 at 50-100 $\%$.  The three time intervals from left to right are 13:58:00-13:59:00 UT, 13:59:00-14:00:00 UT, and 14:03:45-14:04:45 UT.  There were not enough counts at the latest time at 25-50~keV to construct a meaningful image.  Note that the scale is different for every image and decreased with time.}
\label{fig:HXR_images}
\end{figure*}

The X-ray emission at three different time intervals is shown in Figure \ref{fig:HXR_images} at 12-25~keV and 25-50~keV, integrated over a period of 1 minute using the PIXON algorithm \citep{Metcalf_etal1996}.  For the 25-50~keV images we only display between 50-100 $\%$ of the count flux because the flux is so low.  The first time interval in Figure \ref{fig:HXR_images} shows the most intense part of the flare whilst the second and third time interval shows the decay of the flare.  Initially, the X-ray emission is detected on the left-hand side of the active region in both energy ranges.  The rise phase before 13:58:00 UT also shows this behaviour.  The decay of the flare shows a source on the right hand side (west) of the active region but with a much weaker flux.  The source is very extended, having a length of approximately 100 arcsecs from the emission on the left hand side.  There are virtually no 25-50~keV photons in this extended source.  Much later in the flare (e.g. around 14:03:45 UT), the emission is only observed on the right-hand side of the active region at 12-25~keV energies.  The signal is too low above 25~keV to make a meaningful image.  The data after 14:00:00 UT experienced some drop-outs therefore we tried a variety of different imaging algorithms (Clean, Pixon, UVsmooth), which all found a similar structure of the extended source.




\subsection{X-ray lightcurves}

The X-ray lightcurves for the event are given in Figure \ref{fig:hsi_tra_lc}.  The emission starts to rise above the background at 13:57:00 UT although several counts in linear space are not detected until about 13:57:30 UT.  After 13:57:50 UT, attenuator 1 is in operation which absorbs some low energy photons ($<25$~keV) but has no real effect on higher energy photons \citep{Smith_etal2002}.  As such, the peak of the 12-25~keV curve at 13:57:50 UT in Figure \ref{fig:hsi_tra_lc} is artificial.  The attenuator change also caused a brief high-count artefact in the 25-50 keV X-rays at 13:57:50 UT.  We removed this artefact by manually setting 0.5 s of data at this time to zero.  The peak time of the 25-50~keV curve is at 13:58:00 UT.  An additional, smaller peak is seen around 13:58:25 UT.  There are similarly two peaks in the 12-25~keV curve at similar times to the peaks at 25-50~keV.  However, the first peak that occurs around 13:58:00 UT is smaller than the second peak that occurs around 13:58:25 UT for the 12-25 keV energy range.



\section{HXR-UV comparison}

\subsection{HXR-UV morphology}


\begin{figure} \centering
 \includegraphics[width=0.99\columnwidth]{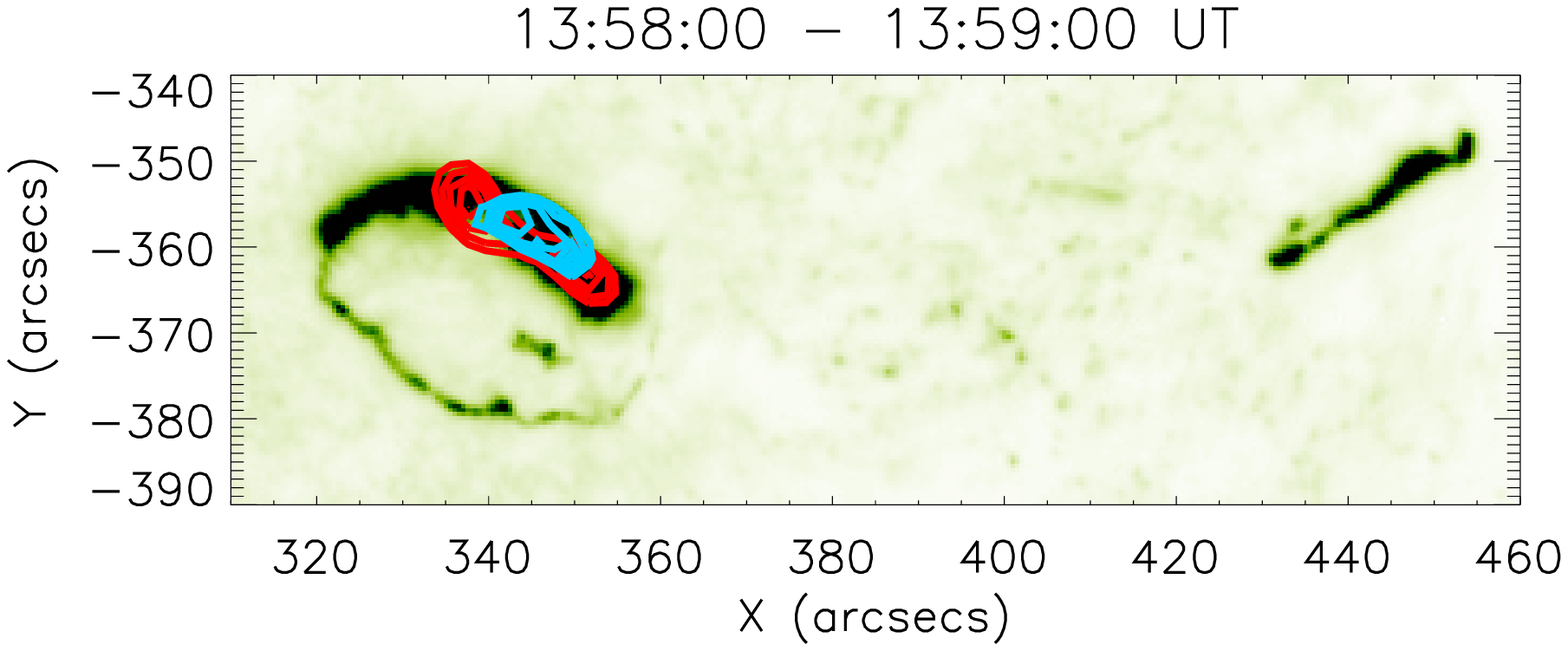}
 \includegraphics[width=0.99\columnwidth]{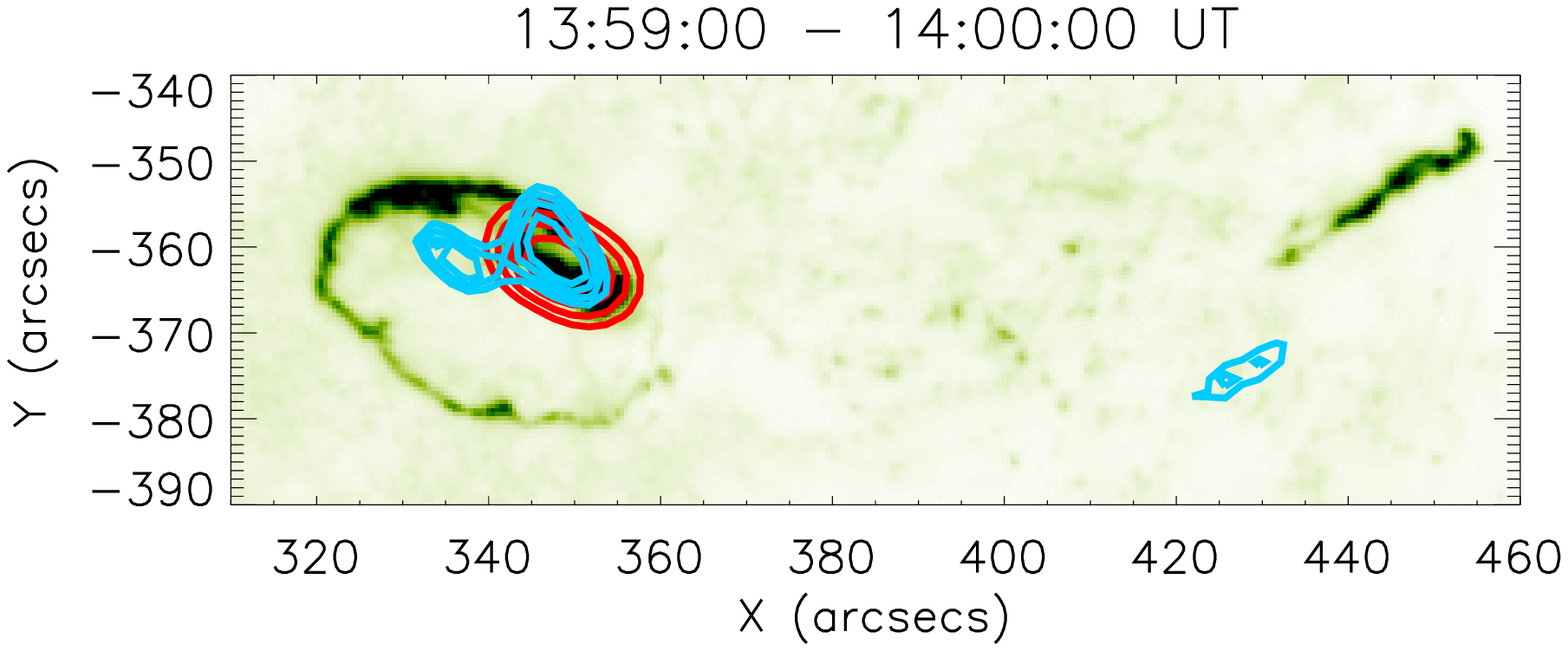}
 \includegraphics[width=0.99\columnwidth]{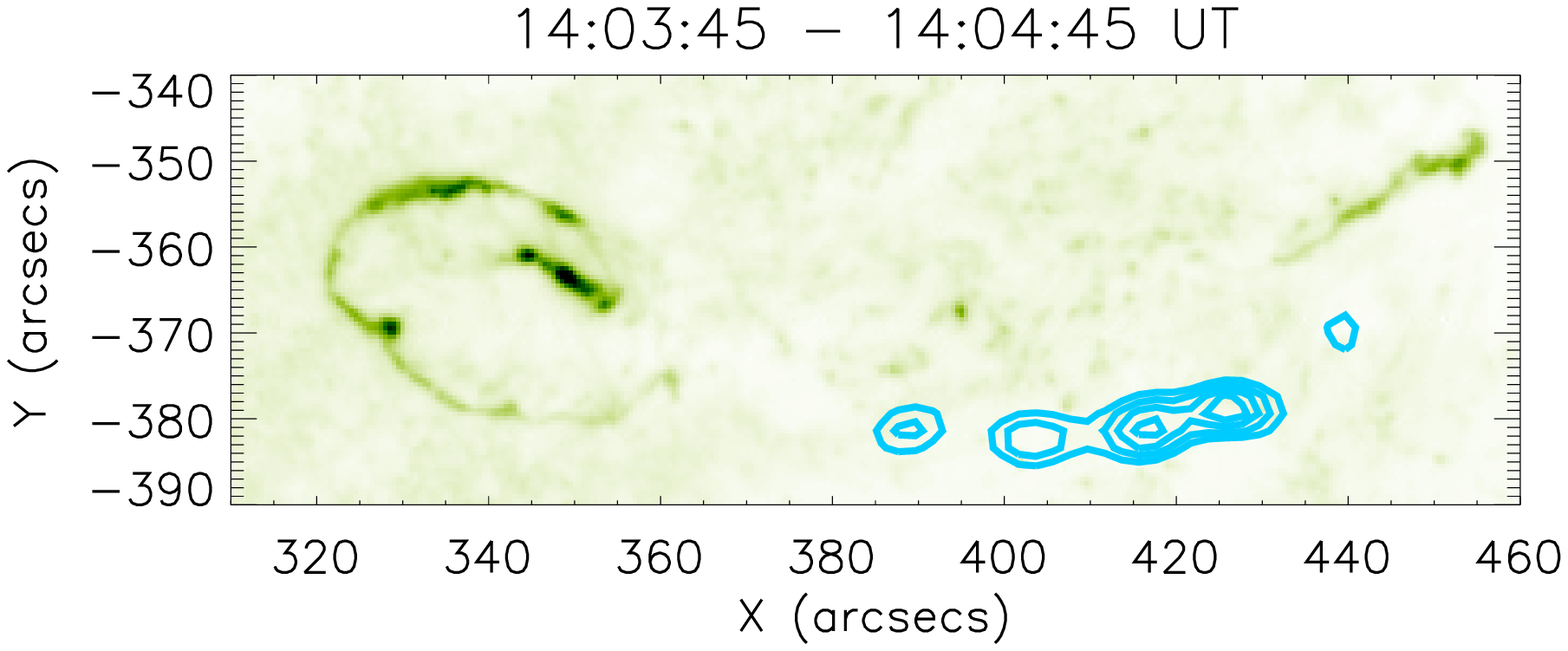}
\caption{RHESSI contours overplotted on a TRACE background.  The RHESSI time intervals are 13:58:00-13:59:00 UT, 13:59:00-14:00:00 UT, 14:03:45-14:04:45 UT.  The contours represent the same data as in Figure \ref{fig:HXR_images}.  The RHESSI contours are 12-25~keV (blue) and 25-50~keV (red) and represent 60-100$\%$ of the maximum amplitude of each time period.  The 1600 $\text{\AA}$ TRACE images are at the central time of the associated RHESSI time integration.}
\label{fig:hsi_tra_morph}
\end{figure}

Figure \ref{fig:hsi_tra_morph} shows the HXR emission from Figure \ref{fig:HXR_images} as contours over the UV emission.  The TRACE pointing was corrected by co-alignment of the TRACE 1600 $\text{\AA}$ image at 14:00:00~UT and the MDI magnetogram at 14:24:00~UT.  The alignment was made manually using the faculae and sunspots with the precise details found in \citet{Aulanier_etal2000}.  We found a shifting of the TRACE image by (4, -3.5) arcsecs provided an alignment of the strong UV quasi-circular ribbon with the polarity inversion line and the shape of the weaker areas of UV emission with the shape of the magnetic field.  The requirement of correcting the TRACE pointing induces an uncertainty into Figure \ref{fig:hsi_tra_morph} of approximately a few arcsecs.  The UV emission, which has a better cadence, is displayed at the central time of the RHESSI image time integration.  

The first image in Figure \ref{fig:hsi_tra_morph} shows the majority of the flare impulsive phase.  The X-rays overlap exactly at the point where the bulk of UV counts are emitted, close to the null point in the magnetic field extrapolation.  We observe what appears to be a flare loop with two non-thermal 25-50 keV footpoints on either side of a thermal 12-25 keV source.  This is reinforced by the apparent curvature in the 12-25 keV source evident from Figure \ref{fig:HXR_images}.  The X-rays are \emph{not} co-spatial with the bulk of the quasi-circular ribbon nor with the ribbon on the right=hand side (RB).  A faint X-ray source could be present but RHESSI does not have the dynamic range to detect faint X-rays in the presence of a strong X-ray source (see Section \ref{ref:disc} for discussion).

Another important observation from Figure \ref{fig:hsi_tra_morph} is that the extended source is \emph{not} co-spatial with \emph{any} of the UV emission.  The elongation of the 12-25~keV X-ray source is approximately 100 arcsecs but it does not reach the UV ribbon on the right side of the active region. 

\begin{figure} \centering
 \includegraphics[width=0.99\columnwidth]{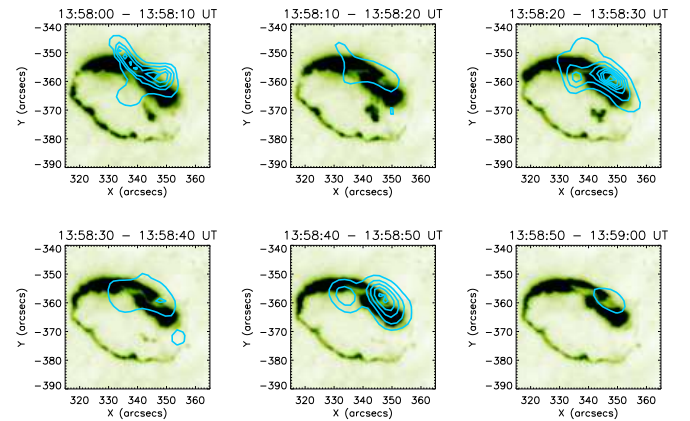}
\caption{RHESSI contours overplotted on a TRACE background for the 60-second period between 13:58:00 and 13:59:00 UT.  The RHESSI images are a 10-second time integration in the energy range 12-25 keV using the PIXON algorithm and detectors 3-8.  The RHESSI contours for all images are at the same level at which the maximum X-ray amplitude is fixed at 4.2 photons cm$^{⁻2}$ s$^{⁻1}$ asec$^{-2}$ and each contour represents 10$\%$ intervals.  The 1600 $\text{\AA}$ TRACE images are at the central time of the associated RHESSI time integration.}
\label{fig:hsi_tra_morph_det}
\end{figure}

Because the majority of the X-ray emission occurs between 13:58:00 UT and 13:59:00 UT, we investigated this period in more detail.  Figure \ref{fig:hsi_tra_morph_det} shows how the 12-25 keV X-ray emission varies as a function of position with respect to the UV emission when we consider a 10-second integration time.  We did not plot the 25-50 keV emission because the number of counts during the 10 second integration times were small.  The contour levels were kept constant in all six images to better display the fluctuating intensity levels of the X-ray emission.  

The main X-ray source remains stationary during the 60-second period centred at roughly (350, -360).  The position lies between the UV ribbons associated with the inner spine and the fan surface but we must be careful to keep in mind the small uncertainty induced by correcting the TRACE pointing.  A secondary X-ray source can be observed on the left-hand side ($x=330$~arcsecs) which displays a general downwards motion during the 60 seconds from $y=-350$~arcsecs to $y=-360$~arcsecs.  There is also a small source observable at 13:58:10-13:58:20 UT and 13:58:30-13:58:40 UT when the main bulk of 12-25~keV emission is weaker.  The X-ray source is quite faint, which could be the reason why it is not observable when there is a high X-ray count rate between 13:58:20-13:58:30 UT.  This source is near the small southern UV source, around (345, -370).  The time range 13:58:10-13:58:20 UT is exactly the range in which the peak of this small UV source occurs (curve B in Figure \ref{fig:hsi_tra_lc}) although it did not display any temporal correlation with the bulk of the X-ray emission.  The small source of both UV and X-ray emission might be related to the reconnecting current sheet around the null point, but the available data cannot prove this conjecture.


\subsection{HXR-UV lightcurves}

There is good co-temporal agreement between the X-ray and UV lightcurves (Figure \ref{fig:hsi_tra_lc}).  The rise time and peak time of area E is very similar to the rise time and first peak time of the X-ray emission in both energy ranges.  Emission starts to significantly rise at roughly 13:57:30 UT and peaks at 13:58:00 UT.  The decay time of the 25-50 keV X-rays and the UV emission is not similar.  This is expected because of the impulsive nature of high-energy X-rays and the long emission time of 1600 $\text{\AA}$ UV light.  We observe a similarly good correlation between the peak time of the right side of the active region in UV (area F) and the second peak of the X-rays in the two energy bands.  The slow decay of the 12-25 keV X-rays mirrors the slow decay of the UV emission in both areas.

Looking at the subregions, the peak in area C has a narrow width in time, similar to the first X-ray peak at 25-50 keV.  The peak in area D occurs with the second peak in X-rays at 13:58:25 UT but we observe an increase in emission that corresponds to the first peak in X-rays.  Area B, the small source, indeed looks anti-correlated in time with the X-ray emission.  However, there were few UV counts for this region and consequently any correlated emission in X-ray would be weak and undetectable with respect to the other X-ray emitting areas.

\begin{figure}\centering
 \includegraphics[width=0.99\columnwidth]{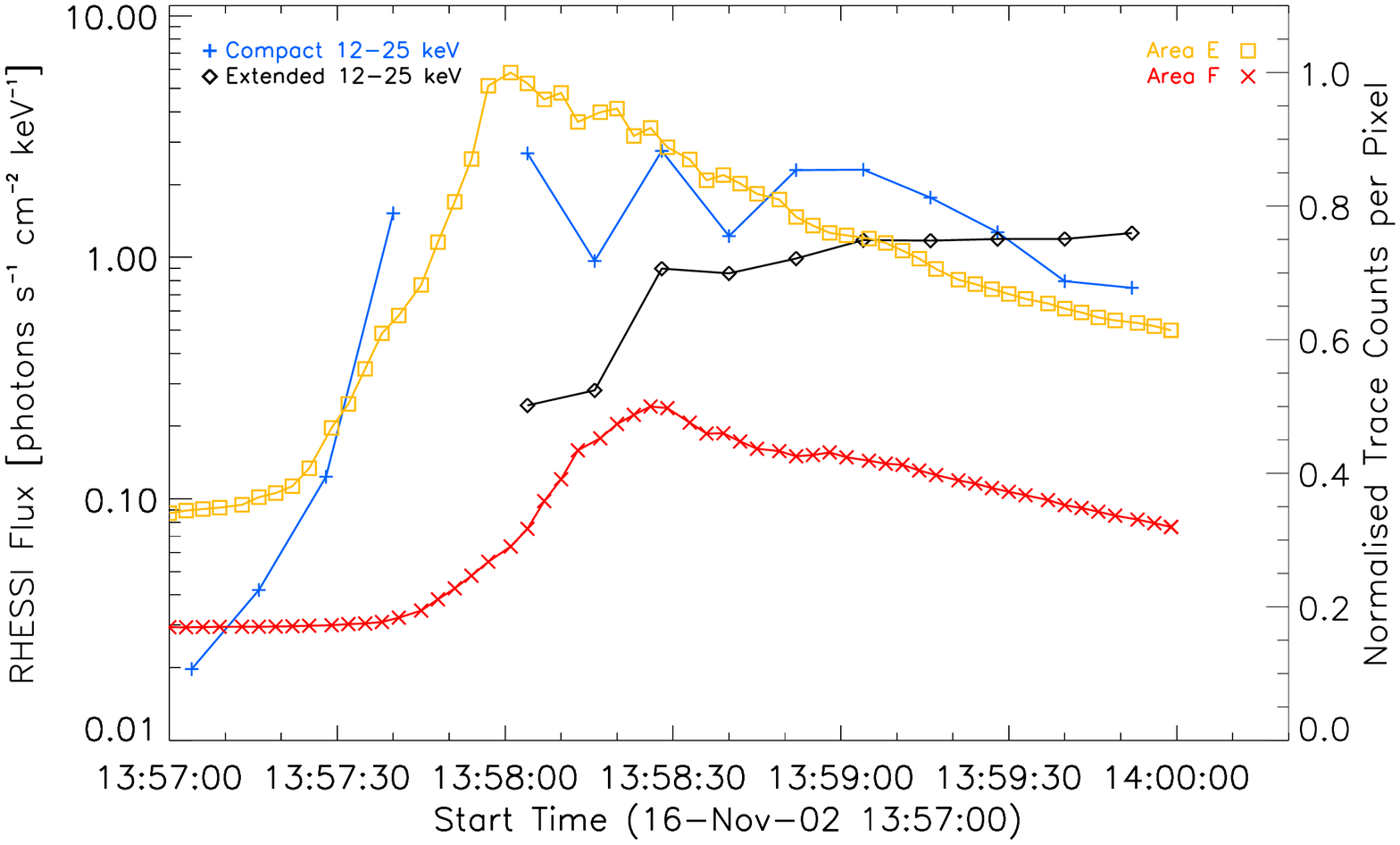}
\caption{RHESSI photon flux for the compact left region and the extended right region with a logarithmic axis for the RHESSI data at 12-25 keV.  The PIXON images used to create the RHESSI light curves had a 12 second integration time using detectors 3-9.  The UV curves for area E, and F are also included with a normalised axis of 1 and 0.5, respectively.}
\label{fig:hsi_tra_lc_reg2}
\end{figure}

To investigate the time dependence of the extended source in X-rays, we created PIXON images between 13:57:00 and 14:00:00 with a 12-second time integration (three RHESSI periods).  We analysed these lightcurves carefully because the time integration is short with respect to the X-ray counts and we assumed that the output of the PIXON algorithm is correct.  We split the flare into two regions, one for the compact X-ray source and one for the extended X-ray source.  The time profile for both sources is displayed in Figure \ref{fig:hsi_tra_lc_reg2} along with the time profile for the UV emission.  We found the extended source starts around 13:58:00 UT with a low level of emission but becomes much stronger around 13:58:30 UT.  The extended source is not co-temporal with the bulk of the UV emission (area E), which is expected.  The growth of the extended X-ray source mirrors the growth of the UV ribbon in area F although the X-ray source continues to increase after 13:58:30 UT.  The extended X-ray source continues to emit at low energies well after the 25-50 keV counts have dwindled down to the background level (Figure \ref{fig:hsi_tra_lc}).

\section{Discussion} \label{ref:disc}

\subsection{Observations}

We have examined the flare of 16 November 2002 flare in the context of X-ray and UV observations.  We found a good temporal and spatial agreement between the X-ray and UV observations during the impulsive phase of the flare.  The spatial location of the X-ray emission was localised in the same location in which we observed the strongest UV emission, which is also directly below the calculated coronal null point in the magnetic field.  This result agrees with other observations \citep[e.g.][]{Temmer_etal2007} that had found that energy release rates are non-uniform along UV flare ribbons with X-ray observations being focussed on areas of the highest energy release rates.  Two peaks of emission were observed, one at 13:58:00 UT associated with the bulk of UV emission and high-energy 25-50 keV X-rays, and the other at 13:58:25 UT associated with a UV ribbon separated from the bulk of the UV emission, the lower energy 12-25 keV X-rays and the emergence of an extended X-ray source.

We saw between 13:58:00 and 13:59:00 UT (Figure \ref{fig:hsi_tra_morph}) that the bulk of the 12-25 keV emission originates between two footpoints of 25-50 keV emission and could be situated in the corona at the top of a loop formed between the fan and the inner spine.  This 12-25 keV emission is thermally distributed and is likely caused by chromospheric evaporation from the accelerated particles at 13:58:00 UT.  Conforming to the Neupert effect, the non-thermal 25-50 keV particles agree reasonably with the time derivative of this 12-25 keV source until 13:59:00 UT.  



During the decay phase of the flare the 25-50~keV X-ray emission ceased but an extended source at $<25$~keV is observed between the two UV ribbons on either side of the active region (Panel 3 in Figure \ref{fig:hsi_tra_morph}).  The extended source of X-ray emission is likely to lie in the corona, on the magnetic loops that connect the two UV ribbons.  The projection effects associated with an active region in the south-west of the Sun and the elevated altitude of the coronal source explain the position of this source.  This is also supported by the apparent loop shape of the X-ray source and the lack of a low-altitude UV counterpart.  Unfortunately, there are no extreme UV (e.g. 195~$\text{\AA}$) observations, which are normally correlated to thermal, coronal X-ray sources.  The extended source implies magnetic connectivity between the two UV ribbons and reinforces the conclusion drawn by MPAS09 that we see a spine/fan magnetic configuration with an outer spine connecting the null point region to the right UV ribbon.  Moreover, it is similar to the late-phase X-ray emission reported by \citet{AlexanderCoyner2006} and \citet{Liu_etal2007} for two separate events.  


The extended X-ray source can be best modelled spectrally by a single thermal distribution with a temperature of about 20~MK (which decreases with time as the plasma cools).  We estimated $\geqq 100$~arcsecs (70~Mm) as the distance that the hot plasma must travel from the area in which we observe the $> 25$ keV X-rays to reach the top of the loop.  This requires at least 70 seconds, even assuming a high velocity of 1000 km s$^{-1}$ for chromospheric evaporation, because hot plasma requires time to fill the coronal loop \citep[e.g.][]{LiGan2006}.  One can see in Figure \ref{fig:hsi_tra_lc_reg2} a 12-25 keV X-ray signature at 13:57:00 UT, which at this time is mostly non-thermal.  It could be responsible for the extended source detected around 13:58:30 UT.  The low level of emission in the extended source at 13:58:00 UT is indeed detected closer than 70~Mm to the main X-ray source.  Chromospheric absorption from the site at which we detected the main X-ray source is the most likely source of the extended X-ray source.  However, because of the ambiguity caused by the high-velocity requirement, we review other possible sources.

One scenario is emission from an electron beam in the corona \citep[e.g.][]{Veronig_etal2005} but no power-law signatures were detected in the spectrum of the extended X-ray source.  Another possible explanation is chromospheric evaporation from the electron beams responsible for the UV ribbon on the right side of the active region (area F in Figure \ref{fig:UV_morph}).  Using the emission measure and temperature from a spectral fit to the large extended X-ray source size we find the energy contained in the hot plasma on the order of $10^{29}$~ergs assuming a volume given by area$^{1.5}$ and a filling factor of 1 \citep{Saint-HilaireBenz2005}.  Given this high-energy requirement, one would expect to observe some X-ray emission co-spatial with the UV ribbon on the right side, where none is detected.  

Thermal conduction from heating at the energy release site has also been invoked to explain some soft X-ray coronal sources in weaker flares \citep[e.g.][and references within]{Veronig_etal2002}.  The thermal conduction front would propagate at the speeds around the ion sound speed of the plasma - in this case approximately 740 km s$^{-1}$ for 20 MK (assuming equal electron and ion temperatures).  We therefore cannot rule out any contribution to the extended X-ray source by thermal conduction but the long loop length of $\geqq 100$~arcsecs would require significant heating very early in the flare.

\subsection{Interpretation}

\begin{figure*}
 \sidecaption
 \includegraphics[width=12cm]{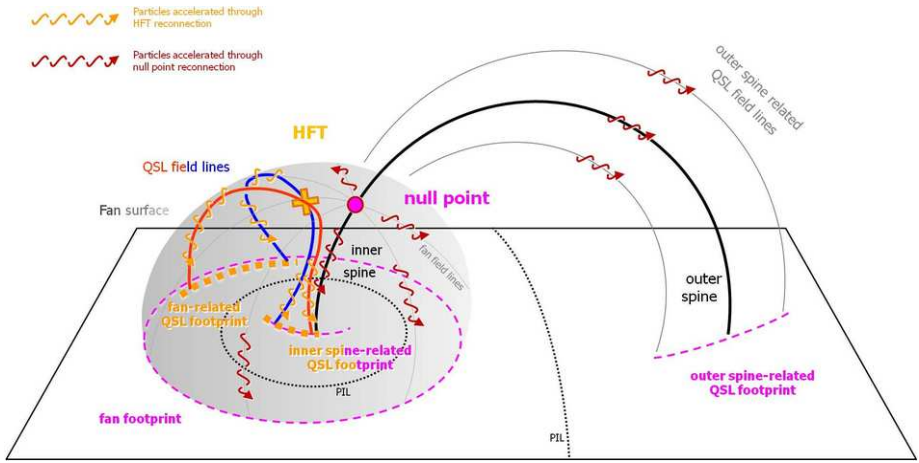}
\caption{Geometry of the magnetic field including the inner/outer spines, the fan surface, and the null point.  The inner field associated with the quasi-separatrix layer are denoted in blue and red.  Particle trajectories from the two different forms of reconnection are shown in brown and orange.}
\label{fig:interp}
\end{figure*}
%

While there is some agreement in space and time between the X-ray and UV observations, there are several intriguing properties that one has to incorporate into any model or explanation of the evolution in the flare.  Specifically, one has to consider the approximate 25 second delay between the peak of the UV emission on the left-hand and right-hand side of the active region, the presence of co-spatial X-rays only where the UV emission is the strongest, and the absence of increased X-ray and UV emission south of the null point.

Our first explanation is based on the properties of null point reconnection that field dispersion increases with distance from the null point.  Given the proximity of the null to the majority of the UV and X-ray emission, we were able to deduce that the energy density of accelerated particles close to the null is higher. Particles that travel farther to the bottom left of the quasi-circular ribbon and to the UV ribbon associated with the outer spine spread much more, which decreases their energy density.  When particles arrive at the chromosphere, they only produce an observable level of X-ray emission where there is a high-energy density.  This explanation can account for why the area emitting the strongest UV emission is the only area in which we observe co-spatial X-rays.  It cannot account for the 25 second delay between the UV emission on the left-hand and right-hand side of the active region, nor does it account for the expected X-ray or UV emission south of the null point.



The second explanation (see Figure \ref{fig:interp} for an overview) puts the emphasis on QSL reconnection, i.e. slip-running reconnection (Aulanier et al. 06) to explain the majority of the UV and X-ray emission.  Magnetically, the UV and X-ray emission occurs at the base of the inner spine and the upper right part of the fan (the cartoon shows it on the left for clarity).  As noted by MPAS09, an extended QSL is present around the null point and is particularly extended around the inner spine (observed in Figure 8 of MPAS09).  We can assume that the point where the inner spine is linked with the upper part of the circular fan (as indicated in orange in Figure \ref{fig:interp}) is energetically very important during the solar flare. This was not specifically observed in MPAS09, who worked with an initially potential field.  It is reasonable to assume, however, that a relatively large amount of free energy and shear is stored in the inner domain of the 3D fan before the eruption \citep[as in the jet model of][]{Pariat_etal2009}.  The simulation of MPAS09 indeed shows that QSL reconnection easily occurs in the inner domain even though the driver was applied in the outer connectivity region.  The strongest currents mostly develop in the core of a QSL, known as a hyperbolic flux tube (HFT) \citep{Titov_etal2002}.

We hypothesise that the first burst of X-rays at 13:58:00 UT and the peak of the UV emission on the left side of the active region is mainly caused by particles that are accelerated through HFT reconnection (the most energetic part of QSL reconnection) from the QSLs related to the inner spine.  With the accelerated particles mainly confined to the inner domain, we can explain why the X-rays are focused at the base of the inner spine and why the peak of UV emission in the separated ribbon on the right hand side of the active region (area F) is not found at 13:58:00 UT.  Moreover, between 13:58:00 and 13:59:00 UT the X-rays take the form of a flare loop, where we observe the thermal 12-25 keV X-rays in between two non-thermal 25-50 keV X-ray footpoints (Figure \ref{fig:hsi_tra_morph}).  This loop could be the tracer from the burst of particles accelerated through HFT reconnection (depicted in Figure \ref{fig:interp} in red and blue).  As noted in \citet{Aulanier_etal2005}, the finite width of QSLs can build up electric currents over a longer time than standard separatrices and hence have a larger capacity for energy storage before the QSL-related current-sheet becomes thin enough to start reconnection. The greater capacity for energy release would correspond to more accelerated particles when the instability occurred and explains the higher intensity of X-ray and UV emission during the first peak of emission at 13:58:00 UT.  The loop-like QSL connectivity denoted in Figure \ref{fig:interp} in red and blue would focus the accelerated particles in a small portion of the fan (top), which can explain the reduced UV emission in the southern part of the quasi-circular ribbon and the lack of detected co-spatial X-ray emission.

Whilst we relate the bulk of the accelerated particles to HFT reconnection, the null point reconnection would still be present.  \citet{Masson_etal2009} observed a bright kernel of emission on the right side of the active region at 13:57:32 UT that develops over time to form the ribbon associated with area F (ribbon RB in MPAS09).  This is an indication that accelerated particles are able to flow from the null point along the outer spine for most of the flare.  Heated plasma must also be able to flow along the outer spine early in the flare to explain the presence of the extended X-ray source.  However, in contrast to the finite width of QSLs, the zero thickness of a null point separatrix causes current sheets to form at the dissipative scale and hence dissipate straight away.  Without a large build-up of energy the flux of accelerated particles will be low.  We suggest that as reconnection develops, the magnetic field lines undergoing QSL/HFT/slip-running reconnection will `slip' towards the null point.  A higher flux of accelerated particles is then able to flow along the outer spine.  The delay between the rise of the UV emission related to the inner and outer domain (this can be seen in Figure \ref{fig:hsi_tra_lc}) is related to the time for the slipping of the magnetic field and is also mirrored in the delay between the peaks of the UV emission.  The majority of energy release is concentrated in the inner domain of the magnetic field and, together with the spreading of the magnetic field in the outer domain, can explain the absence of detectable X-ray emission above 25 keV at the base of the outer spine (co-spatial with the ribbon in area F in Figure \ref{fig:UV_morph}).

Whilst this second scenario is attractive because it explains the delays and difference in energy deposition, it is based on the assumption that an important HFT is present in the inner domain.  The scenario cannot be completely proven, but no arguments preclude its existence either.  If proven, this scenario implies that HFT reconnection could be a major driver of energy release in solar flares even when a true null point is involved.  This statement, while far from being fully demonstrated, is sustained by the capacity of QSL to carry more intense currents than separatrices and that the shape and morphology of the ribbons in the flare on the 16th November 2002 is governed by the QSLs and not only by the null point topology.  The larger area of QSLs is also advantageous for explaining the high numbers of accelerated electrons that is required to explain many non-thermal X-ray observations.  More studies that not only consider the X-ray and UV emission but are able to constrain and analyse the structure of the magnetic field would be advantageous for continuing to assess the applicability of HFT reconnection in solar flares.

\begin{acknowledgements}
Financial support from the SOLAIRE Network (MTRN-CT-2006-035484), the HESPE Network (FP7-SPACE-2010-263086) and the PHC Alliance Programme between France and the UK are gratefully acknowledged.  We also acknowledge support from the Centre National d'Etudes Spatiales (CNES) and from the French program on Solar-Terrestrial Physics (PNST) of INSU/CNRS for the participation to the RHESSI project.  We thank S. Masson for useful comments and advice.  We also thank the anonymous referee for useful comments.
\end{acknowledgements}

\bibliographystyle{aa}
\bibliography{hxr_uv}

\end{document}